\documentclass[prd,aps,twocolumn,amsmath,amssymb,nofootinbib,preprintnumbers]
{revtex4}

\usepackage{graphicx}
\usepackage{dcolumn}
\usepackage{bm}
\usepackage{amsmath}
\usepackage{amsthm}

\usepackage{amsfonts}
\usepackage{euscript,bbm}
\usepackage{ifthen}
\usepackage{psfrag}
\usepackage{slashed}

\def\ls{\mathrel{\lower4pt\vbox{\lineskip=0pt\baselineskip=0pt
           \hbox{$<$}\hbox{$\sim$}}}}
\def\gs{\mathrel{\lower4pt\vbox{\lineskip=0pt\baselineskip=0pt
           \hbox{$>$}\hbox{$\sim$}}}}
\def\drawbox#1#2{\hrule height#2pt

\hbox{\vrule width#2pt height#1pt \kern#1pt
              \vrule width#2pt}
              \hrule height#2pt}

\def\Asym#1#2{\vcenter{\vbox{\drawbox{#1}{#2}
              \kern-#2pt       
              \drawbox{#1}{#2}}}}

\newcommand{\be}{\begin{equation}}
\newcommand{\ee}{\end{equation}}
\newcommand{\bea}{\begin{eqnarray}}
\newcommand{\eea}{\end{eqnarray}}

\newcommand{\lsim}{\mathrel{\hbox{\rlap{\lower.55ex\hbox{$\sim$}} \kern-.3em \raise.4ex \hbox{$<$}}}}
\newcommand{\gsim}{\mathrel{\hbox{\rlap{\lower.55ex\hbox{$\sim$}} \kern-.3em \raise.4ex \hbox{$>$}}}}
\newcommand{\beq}{\begin{equation}}
\newcommand{\eeq}{\end{equation}}
\newcommand{\drm}{\mathrm{d}}

\newcommand{\krh}{k_\mathrm{RH}}

\newcommand{\sigv}{\langle \sigma v \rangle}
\newcommand{\gstarRH}{g_{*\mathrm{RH}}}
\newcommand{\Trh}{T_\mathrm{RH}}

\newcommand{\mdm}{m_\chi}
\newcommand{\sigunits}{\,\mathrm{cm^3\,s^{-1}}}
\newcommand{\kcut}{k_\mathrm{cut}}

\newcommand{\Tkd}{T_\mathrm{kd}}
\newcommand{\kkd}{k_\mathrm{kd}}
\newcommand{\kfs}{k_\mathrm{fs}}

\begin{document}

\title{Bringing Isolated Dark Matter Out of Isolation: \\
Late-time Reheating and Indirect Detection}

\author{Adrienne L. Erickcek$^{1}$}
\author{Kuver Sinha$^{2}$}
\author{Scott Watson$^{2}$}

\affiliation{$^{1}$~Department of Physics and Astronomy, University of North Carolina at Chapel Hill,\\
Phillips Hall CB 3255, Chapel Hill, NC 27599 USA \\
$^{2}$~Department of Physics, Syracuse University, Syracuse, NY 13244, USA 
}

\date{\today}

\begin{abstract}

In standard cosmology, the growth of structure becomes significant following matter-radiation equality. In non-thermal histories, where an effectively matter-dominated phase occurs due to scalar oscillations prior to Big Bang Nucleosynthesis, a new scale at smaller wavelengths appears in the matter power spectrum. Density perturbations that enter the horizon during the matter-dominated phase grow linearly with the scale factor prior to the onset of radiation domination, which leads to enhanced inhomogeneity on small scales if dark matter thermally and kinetically
decouples during the matter-dominated phase. The microhalos that form from these enhanced perturbations significantly boost the self-annihilation rate for dark matter. This has important implications for indirect detection experiments: the larger annihilation rate will result in observable signals from dark matter candidates that are usually deemed untestable. As a proof of principle, we consider Binos in heavy supersymmetry with an intermediate extended Higgs sector and all other superpartners decoupled. We find that these isolated Binos, which lie under the neutrino floor, can account for the dark matter relic density while also leading to observable predictions for \textit{Fermi}-LAT. Current limits on the annihilation cross section from \textit{Fermi}-LAT's observations of
dwarf spheroidal galaxies may already constrain Bino dark matter up to masses $\mathcal{O}(300)$ GeV, depending on the internal structure of the microhalos.  More extensive constraints are possible with improved gamma-ray bounds and boost calculations from $N$-body simulations.

\end{abstract}

\maketitle


{\it {\bf Introduction.}} One missing piece in our reconstruction of the history of the Universe is the period between inflation and the onset of Big Bang Nucleosynthesis (BBN). 
When computing the dark matter (DM) abundance predicted by a particular extension of the standard model, it is customary to assume that the Universe was radiation dominated long before BBN.  However, deviations from radiation domination in the early Universe are required in order to provide the primordial perturbations necessary to seed the growth of large-scale structure, with cosmic inflation providing a compelling explanation. This raises the question, when did the Universe become radiation dominated?  Both prolonged inflationary reheating and the existence of gravitationally coupled scalars (moduli) provide independent motivation that the Universe could have been matter dominated until the time of BBN.  In both situations, oscillating scalar fields dominate the energy density of the Universe, leading to an Early Matter-Dominated Epoch (EMDE) prior to BBN \cite{Kane:2015jia}.   If DM thermally decouples during an EMDE, the relationship between its annihilation cross section and its current abundance radically changes, and particle physics models that predict too much DM in standard thermal histories become viable \cite{Giudice:2000ex}, \cite{Gelmini:2006pw}.

The impact of an EMDE on the evolution of small-scale structure provides the means to constrain these scenarios \cite{Erickcek:2011us,Fan:2014zua, Erickcek:2015jza}. The key point is that while matter perturbations only grow logarithmically with the scale factor during radiation domination, they grow linearly during an EMDE.  Consequently, perturbations that enter the horizon during the EMDE experience an early stage of linear growth.  If DM decouples both thermally and kinetically prior to the onset of radiation domination, this enhancement of the small-scale matter power spectrum leads to the formation of sub-earth-mass microhalos that contain most of the DM at high redshift.  These microhalos are then the building blocks of subsequent structure formation, and their presence in present-day halos enhances the DM annihilation rate by several orders of magnitude \cite{Erickcek:2015jza}.

In this paper we point out that the boost to the annihilation rate can have profound consequences for models of DM in particle physics,
bringing hitherto untestable DM candidates squarely within the realm of observation. The candidates we consider have $(i)$ annihilation cross sections  that are suppressed by $\mathcal{O}(10^{-3} - 10^{-6})$ compared to the canonical cross section for thermal WIMPs, and $(ii)$ scattering cross sections with atomic nuclei that are typically under the neutrino floor. In standard thermal cosmology, an ``isolated" DM candidate like this overcloses the Universe, and its annihilation rate is too feeble to be constrained by experiment.  In the non-thermal cosmology described above, such a candidate can become acceptable if its number density is diluted by entropy production during the EMDE, and the boosted annihilation rate from microhalos makes it potentially observable. Moreover, the feeble scattering cross section with nuclei ensures that the DM kinetically decouples early enough for the microhalos to survive erasure. Therefore, the methods described in this paper are the only way to constrain such classes of isolated DM.

\begin{figure*}[!t]
\centering
\mbox{\includegraphics[width=8cm,height=8cm,keepaspectratio]{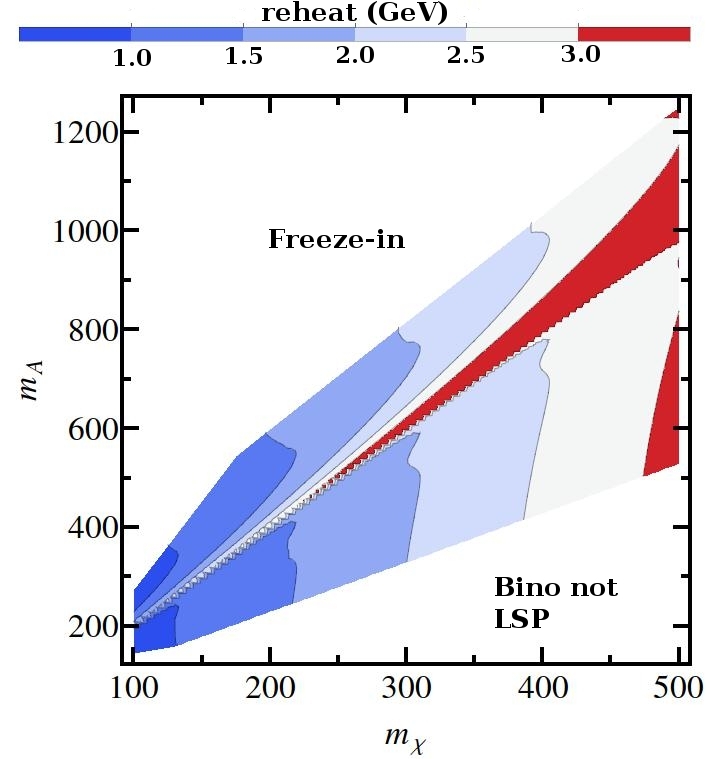}\quad\includegraphics[width=8cm,height=8cm,keepaspectratio]{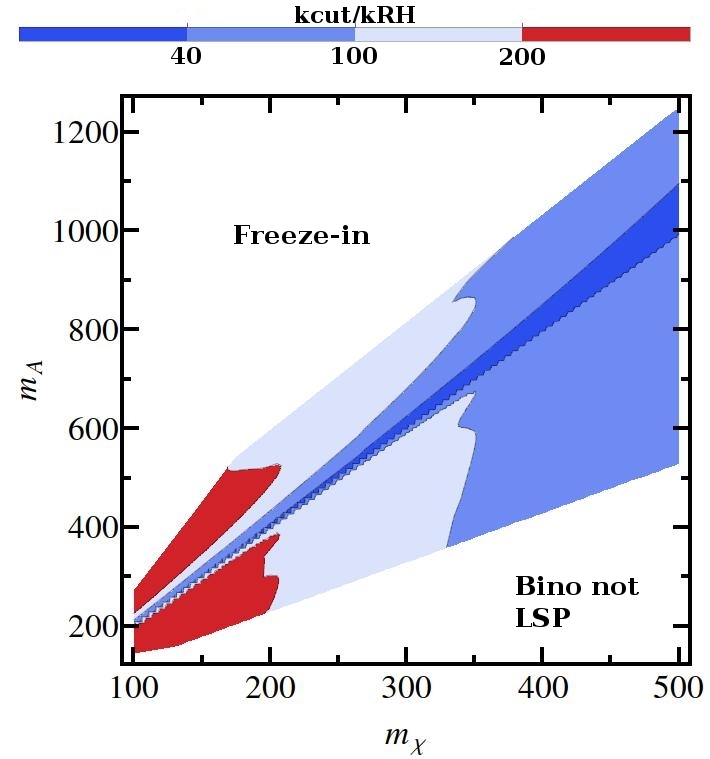}}
\caption{The distribution of reheat temperatures (left panel) and $\kcut / \krh$ (right panel) on the $(m_{\chi}, m_A)$ plane. Here $\kcut = {\rm min}(\kkd, \kfs )$. All points satisfy the relic density constraint. Fixed values of sfermion masses $m_{\tilde{f}} = 60$ TeV, Higgsinos $\mu = 150$ TeV, and $\tan{\beta} = 8$ are assumed.}
\label{Graph1}
\end{figure*}

Any DM model that satisfies the criteria on the kinetic decoupling temperature and annihilation cross section spelled out in Eq.~\eqref{TkdoverTrh} and Eq.~\eqref{sigvrange} can be constrained in this way, but as a proof of principle, we will focus on supersymmetric Bino DM $\chi$ with mass $m_\chi \sim \mathcal{O}(100-500)$ GeV with sfermions, gauginos, and Higgsinos decoupled.
To obtain boosted annihilation rates that are within the realm of observation by \emph{Fermi}-LAT, we will consider models with an intermediate extended Higgs sector. The pseudoscalar Higgs $A$ has mass $m_A \, \sim \, \mathcal{O}(m_\chi )$ - $\mathcal{O}(1200)$ GeV, and the Bino DM annihilates mainly through the $s$-channel. Spectra with heavy supersymmetry and an intermediate Higgs sector have been studied recently \cite{Lee:2015uza}. Apart from the usual reasons to study split supersymmetry, these scenarios are increasingly motivated by bounds on gluinos and charginos coming from the LHC. 

{\it {\bf DM during an EMDE.}}  During the EMDE, we assume that the energy density of the Universe is dominated by an oscillating scalar field that decays into relativistic particles.  The scalar's decay rate determines the reheat temperature $T_\mathrm{RH}$, which is the temperature of the radiation bath when the Universe became radiation dominated \cite[e.g.][]{Giudice:2000ex}.  We also assume that DM thermally decouples (freezes out) during the EMDE and that there is negligible production of DM from scalar decays.  The former assumption requires that the annihilation cross section is large enough to bring DM into thermal equilibrium, failing which it will ``freeze-in", an option we do not consider further because the relevant annihilation cross sections are too small to generate observable signatures \cite{Erickcek:2015jza}.  The latter assumption depends on details of the inflaton or modulus sector, and can be realized if the field couples weakly to $R$-odd particles. 

After DM thermally decouples at a temperature $T_f$, the comoving number density of DM particles remains constant.  In contrast, relativistic particles are continuously created by inflaton/moduli decays during the EMDE, so the DM-to-photon ratio is diluted, and the current DM density is suppressed.  If DM freezes out after reheating, $\sigv \simeq 3 \times 10^{-26} \sigunits$ results in the observed DM density ($\Omega_\chi h^2 = 0.12$, where $\Omega_\chi$ is $\rho_\chi$ divided by the critical density with $H_0 = 100h$ km/s/Mpc \cite{Ade:2015xua}), but if $T_f>T_{{\rm RH}}$ then the resulting DM density is \cite{Giudice:2000ex, Erickcek:2015jza}
\begin{align}
\Omega_\chi h^2 \simeq \,\,& 1.6 \times 10^{-4}\, \frac{\sqrt{g_*(T_{\rm RH})}}{g_*(T_f)} \left( \frac{m_\chi /T_f}{15}\right)^4\left( \frac{150}{m_\chi /T_{\rm RH}}\right)^3 \nonumber \\
&\times\left(\frac{3 \times 10^{-26} \sigunits}{\sigv}\right). \label{OmegaFO}
\end{align}
Therefore, $\sigv \ll 3 \times 10^{-26} \sigunits$ is required to generate the observed DM relic density during an EMDE, and models that would otherwise overclose the Universe become viable. 

After DM thermally decouples, DM is still kept in local kinetic equilibrium by scattering processes with Standard Model (SM) particles.  At temperatures greater than the kinetic decoupling temperature $\Tkd$, which is approximately defined as the temperature at which the 
momentum transfer rate falls below the Hubble expansion rate, DM particles are tightly coupled to the thermal bath, which alters the evolution of the DM density perturbations.  Furthermore, the kinetic energy of the DM particles when they decouple determines their comoving free-streaming horizon
%
$\lambda_\mathrm{fs} = \int^{t_0}_{t_\mathrm{kd}} \frac{v}{a} \drm t \simeq \sqrt{\frac{\Tkd}{\mdm}} a(\Tkd) \int_{a(\Tkd)}^{1} \frac{\drm a}{a^3 H(a)}$.
%
Perturbations with wavelengths smaller than $\lambda_\mathrm{fs}$ are erased, which prevents the formation of microhalos if $\Tkd \lsim T_\mathrm{RH}$.



We will be interested in kinetic decoupling temperatures that are higher than the reheat temperature so that DM kinetically decouples during the EMDE.  Since the expansion rate at a given temperature is faster during the EMDE than during radiation domination, DM decouples at a higher temperature ($\Tkd$) than it would have in a purely radiation dominated era ($T_{\mathrm{kd,RD}}$).  The corrected temperature can be obtained by noting that the elastic scattering cross section of DM ($\sigma_{{\rm el}}$) is proportional to $T^2$, implying that DM decouples when $T_{{\rm kd}} \sim T^2_{{\rm kd,RD}}/T_{{\rm RH}}$ \cite{Gelmini:2008sh}; the exact dependence is given by  \cite{Erickcek:2015jza}:
\beq
T_{{\rm kd}} = \left [ \frac{g_*(T_{{\rm kd}})^2}{g_*(T_{{\rm kd,RD}})\gstarRH}\right]^{1/4}\sqrt{\frac52} \frac{T_{{\rm kd,RD}}^2}{\Trh}.
\eeq

Since our primary example will be Bino DM in supersymmetric models, we give some details about its interactions with SM fermions, which will determine its scattering cross section and decoupling temperature. The scattering cross section of Bino DM with SM particles has been studied in detail by \cite{Hofmann:2001bi}, and implemented in DarkSUSY \cite{Gondolo:2004sc}. 
The scattering cross section between neutralinos and SM fermions is mediated by the exchange of sfermions in the s- and u- channels, and $Z$ and scalar and pseudoscalar Higgs exchange in the t-channel. The coupling of Binos to sfermions and SM fermions can be written as $\mathcal{L} \, = \, -\sqrt{2}g \overline{f} \{\alpha \tilde{f}_L \mathcal{P}_R - \beta \tilde{f}_R  \mathcal{P}_L \} \chi \, + \, h.c.$ where $\alpha = \frac{Y_f}{2} \tan{\theta_W}$ and $\beta = Q_f \tan{\theta_W}$, with $g, Y_f,$ and $Q_f$ being the electroweak coupling constant, weak hypercharge, and the electric charge of the fermion, respectively. We will consider fermion energies $\omega$ in the regime of low momentum transfer, with $\omega \ll m_\chi$, $t \rightarrow 0$, and $s \rightarrow m^2_\chi + 2 m_\chi \omega + m^2_\ell$, where $s$ and $t$ are the usual Mandelstam variables. Note that the approximation $v \sim 1$ for the Moeller velocity is very good in this case. The elastic scattering rate for $\chi + \ell \rightarrow \chi + \ell$, $\Gamma_{\rm el} = \sum_i \left\langle v \sigma_{\rm el} (\omega_\ell) \right\rangle (T) \; n_\ell (T)$ is then given by 
\begin{eqnarray}
\label{elscatt1}
\Gamma_{\rm el} =
\frac{288}{\pi} \; \sum_L (\alpha^{ 4} + \beta ^{4})
\left(
\frac{G_{\rm F} M_W^{\; \; 2}} {m^2_{\tilde{\ell}} - m^2_{\chi} } \right)^2 T^2 \; n_\ell \; ,
\label{elscatt2}
\end{eqnarray}
where $n_\ell$ denotes the number density of leptons. The relaxation time $\tau$ (the time DM needs to return to local thermal equilibrium)  can be determined from the number of scatterings required to change the momentum of a DM particle significantly: $\tau (T) \simeq  \sqrt{2/3} (m_\chi /T) (1/ \Gamma_\mathrm{el})$  \cite{Hofmann:2001bi}. The kinetic decoupling temperature is defined by requiring $\tau \sim 1/H$. For sleptons in the 10-100 TeV mass range, one can obtain $T_{\mathrm{kd,RD}}$ values as high as $\mathcal{O}(1-5)$ GeV.


{\it {\bf Results.}} Obtaining observable enhancement of the DM annihilation rate hinges on two requirements: $(i)$ the EMDE sufficiently enhances small-scale perturbations (the halo is clumpy enough) to greatly boost the DM annihilation rate, and $(ii)$ the annihilation cross section is large enough that the boosted rate falls within the observable range of current and future experiments.  To quantify these two requirements, we define the quantities $\kkd$ and $\krh$, which are the wavenumbers of the modes that enter the horizon at $T = T_{{\rm kd}}$ and $T = T_{{\rm RH}}$, respectively, as well as $\kfs = \lambda^{-1}_\mathrm{fs}$ and $\kcut = {\rm min}(\kkd, \kfs )$; perturbations are exponentially suppressed for $k > \kcut$.

\begin{figure*}[!t]
\centering
\mbox{\includegraphics[width=8cm,height=8cm,keepaspectratio]{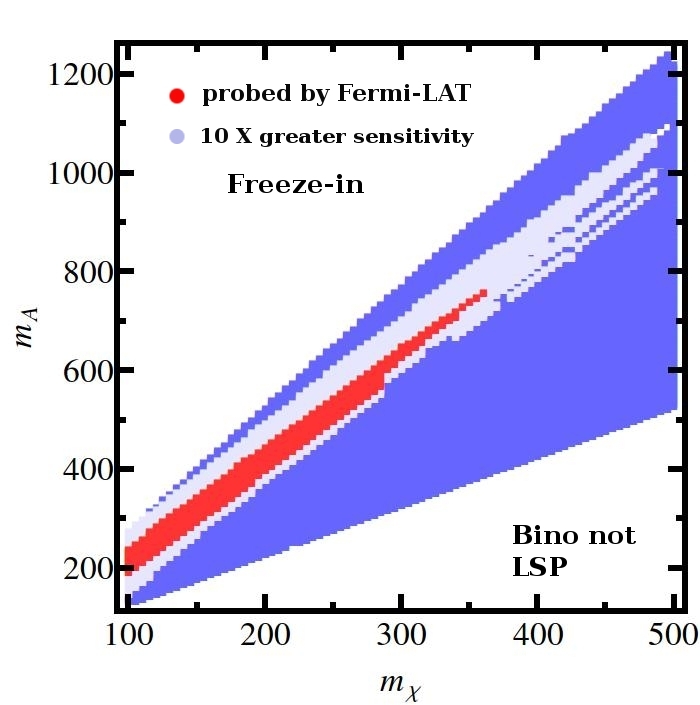}\quad\includegraphics[width=8cm,height=8cm,keepaspectratio]{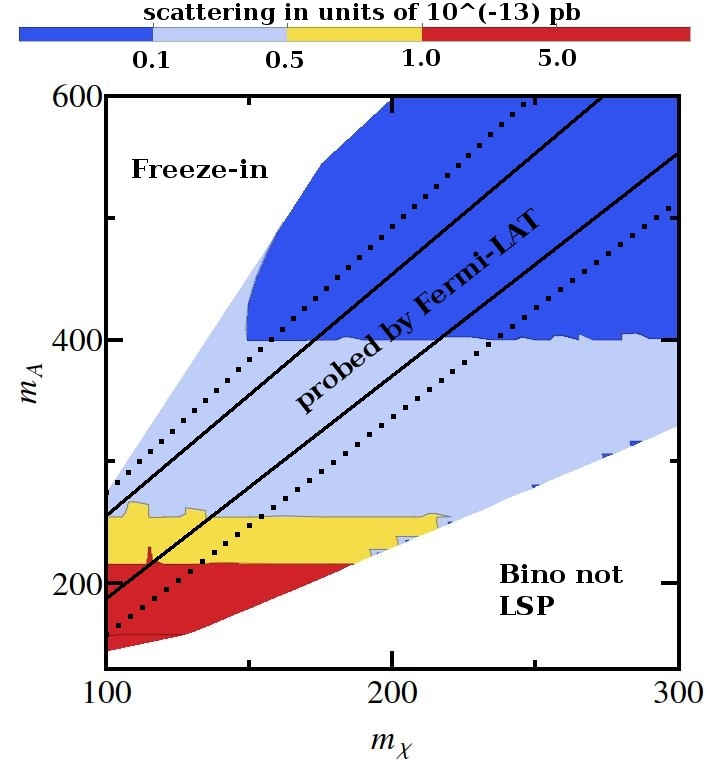}}
\caption{Models constrained by 6-yr Pass 8 \textit{Fermi}-LAT dwarf galaxy data (left panel) and the corresponding scattering cross sections off nuclei in units of $10^{-13}$ pb (right panel). Left panel: The red region is 
ruled out by current observations using optimistic estimates of the EMDE boost factor (see text);
the light blue region is what would be constrained given an $\mathcal{O}(10)$ increase 
in either the sensitivity or the boost factor.
Right panel: Regions that are not red are mostly under the neutrino floor. The solid (dashed) black line shows the region corresponding to red (light blue) of the left panel.}
\label{Graph2}
\end{figure*}

The Press-Schechter formalism \cite{Press:1973iz} requires that the rms density perturbation exceed a critical value for the formation of microhalos.   Therefore, the microhalo population depends strongly on the ratio $\kcut / \krh$, which determines the masses of the smallest microhalos and the timing of their formation.  By studying the dependence of the Press-Schechter differential bound fraction on $\kcut / \krh$, \cite{Erickcek:2015jza} found that the 
EMDE dramatically increases the microhalo population if $\kcut / \krh \, \geq \, 10$, which translates to 
\be \label{TkdoverTrh}
T_{\mathrm{kd,RD}}/T_{{\rm RH}} \, \geq \, 2 
\ee
for values of $T_{{\rm RH}}$ and $m_{\chi}$ in the range we are interested in. This is the first requirement on our supersymmetric parameter space, and can be recast in several other ways. From the relationship between $T_{\mathrm{kd,RD}}$ and $T_{{\rm kd}}$, this translates approximately to $T_{{\rm kd}}/T_{{\rm RH}} > 2\sqrt{10}$, which, since we will require that DM freezes out prior to kinetic decoupling, implies that $T_{f}/T_{{\rm RH}} > 2\sqrt{10}$. 
Since the freeze-out temperature is approximately given as $T_f \sim m_\chi/10$, it turns out that one typically needs $m_\chi/T_{{\rm RH}} > 100 $ for Eq.~\eqref{TkdoverTrh} to hold. Given this range of $m_\chi/T_{{\rm RH}}$, using Eq.~\eqref{OmegaFO}, one obtains that the annihilation cross section is bounded from above. This upper bound is most conveniently expressed as $ \frac{\sigv}{m^2_\chi} \lsim 10^{-16}$ GeV$^{-4}$, as can be checked by the exact numerical calculations. Consequently, we are most interested in particles with
\be \label{sigvrange}
10^{-20} \, {\rm GeV}^{-4} \, \lsim  \,  \frac{\sigv}{m^2_\chi} \, \lsim \, 10^{-16} \, {\rm GeV}^{-4} \,\,,
\ee
where the lower bound arises so that the final boosted annihilation rate is in the realm of current observations and the upper bound ensures that the EMDE enhances the abundance of microhalos.
 
We scan the parameter space of Bino DM in the MSSM, keeping the constraints of Eq.~\eqref{TkdoverTrh} and Eq.~\eqref{sigvrange} in mind. For any given values of  $\sigv$ and $m_\chi$, the reheat temperature $T_{{\rm RH}}$ is chosen such that the relic density constraint $\Omega h^2 = 0.12$ is satisfied from Eq.~\eqref{OmegaFO}.  In order to satisfy Eq.~\eqref{TkdoverTrh}, all sfermions are kept at $\sim \mathcal{O}(60)$ TeV, and charged and neutral Higgsinos at $\mathcal{O}(100)$ TeV. The Winos and the gluino are kept at a few TeV and we choose $\tan{\beta} = 8$. Our results are largely insensitive to changes in their values. The results of the scan are displayed in Fig.~\ref{Graph1}. All models in the scan satisfy the relic density constraint. After obtaining $T_{{\rm kd, RD}}$ from DarkSUSY, we calculate $T_{{\rm kd}}$  and $\lambda_{{\rm fs}}$ , which enables us to find $\kcut/\krh$. The corridor along which the pseudoscalar resonance is most effective corresponds to the highest $T_{{\rm RH}}$ and lowest $\kcut / \krh$ values, as expected. The left panel shows that most models in the scan satisfy $m_\chi/T_{{\rm RH}} > 100 $. 

We now turn to the boost-factor calculation, which will enable us to constrain these models. Using the Press-Schechter formalism to predict the microhalo abundance, \cite{Erickcek:2015jza} estimated the resulting boost to the annihilation rate by assuming that all microhalos present at a certain redshift have NFW profiles with $c = 2$ and that the central regions of these microhalos survive to the present day.  The resulting boost factors are highly sensitive to the redshift $z_f$ at which the microhalo population is evaluated because earlier-forming microhalos have denser central regions.  Since $\kcut/\krh$ determines the redshift at which the first microhalos form, it provides an upper bound on $z_f$: $z_f \lsim 400$ for $40 \gsim \kcut/\krh \gsim 20$ and $z_f \lsim 50$ for $\kcut/\krh \simeq 10$.  We also require that $z_f \gsim 50$ because Press-Schechter microhalo abundance decreases at lower redshifts as larger halos absorb the microhalos.  The appropriate choice of $z_f$ for $\kcut/\krh \gsim 20$ depends on the outcome of microhalo-microhalo mergers: do the microhalos present at $z\simeq 400$ survive as subhalos of the microhalos that contain most of the dark matter at $z\simeq 50$?  If they survive, then $z_f = 400$ provides the best estimate of the boost factor $B$, and for the nearest dwarf spheroidals (dSphs), $1+B \, \simeq \, 20,000$ for $\kcut/\krh \simeq 20$ and $200,000$ for $\kcut/\krh \simeq 40$.  If they are destroyed, then taking $z_f = 50$ reduces the dSph boost factors to between 200 and 500 for $10 \lsim \kcut/\krh \lsim 40$.  We will assume the more optimistic scenario ($z_f \simeq 400$) in this analysis, noting that numerical simulations of microhalo formation in EMDE scenarios are required to robustly determine the appropriate boost factor.   \textit{Fermi}-LAT's constraints on DM annihilation within dSphs are derived assuming that the potential annihilation signal is confined to the dSphs’ central regions.  To account for the fact that this assumption neglects emission from microhalos outside the central region, we reduce the boost factor by a factor of ten.  Therefore, the effective boost factor for dSphs will be taken to be 2,000 and 20,000, for $20 <\kcut/\krh< 40$ and $\kcut/\krh \, > \, 40$, respectively.

With these boost factors, the parameter space that is currently constrained by the \textit{Fermi}-LAT collaboration's 6-year Pass 8 limits on the DM annihilation cross section from dwarf galaxies \cite{Ackermann:2015zua} is shown on the left panel of Fig.~\ref{Graph2}. The red region corresponds to the models that are currently ruled out assuming the boost factors described above; the light blue region is what would be constrained given an $\mathcal{O}(10)$ increase in the sensitivity or the boost factor. The left panel shows the corresponding scattering cross sections relevant for direct detection; regions that are not red are under the neutrino floor. The solid (dashed) black line shows the region corresponding to red (light blue) of the left panel. Since $\tan{\beta} = 8$, there are no LHC constraints coming from $A \rightarrow \tau \tau$ on the plane. Our work is thus currently the only way to probe these regions.

{\it {\bf Outlook.}} DM models that satisfy Eq.~\eqref{TkdoverTrh} and Eq.~\eqref{sigvrange} can be constrained by our methods. These are candidates that are disallowed by the relic density constraint in a thermal history and have scattering cross sections that are too feeble for direct detection. The main example we have investigated is Bino DM in models with heavy supersymmetry and an intermediate Higgs sector.

Depending on the boost factor, Binos in the range $\lsim \mathcal{O}(300)$ GeV are already being probed by \textit{Fermi}-LAT data. The primary source of uncertainty for the boost factor is the internal structure and substructure of the microhalos, and for this, detailed $N$-body simulations are required. A 10x increase in either the \textit{Fermi}-LAT sensitivity or the boost factor would probe an even larger region of parameter space. From the right panel of Fig.~\ref{Graph1}, it is clear that much of the unconstrained region has $\kcut /\krh \, > \, 40$.   A larger value of $\kcut /\krh $ means that microhalos can form much earlier than $z=400$.  Earlier-forming microhalos are denser and thus have larger boost factors, but they also experience more microhalo-microhalo interactions, and it is possibly more difficult for them to survive. While we have conservatively assumed that these microhalos have the same boost factors as for cases with $\kcut /\krh \simeq 40$, the fate of microhalos with large $\kcut$ is an interesting question. On the particle physics side, it would be interesting to explore, in greater detail, supersymmetric spectra with an intermediate extended Higgs sector in light of these DM constraints.

{\it {\bf Acknowledgements.}} We thank JiJi Fan, and Tim Tait for useful comments on an earlier draft of this paper. A.E. is supported by NSF Grant No. PHY-1417446.  K.S. and S.W are supported by NASA Astrophysics Theory Grant NNH12ZDA001N.
S.W. is also supported by DOE grant DE-FG02-85ER40237.


\begin{thebibliography}{99}

\bibitem{Kane:2015jia} 
  G.~Kane, K.~Sinha and S.~Watson,
  Int.\ J.\ Mod.\ Phys.\ D {\bf 24}, no. 08, 1530022 (2015)
  [arXiv:1502.07746 [hep-th]].
  

\bibitem{Giudice:2000ex} 
  G.~F.~Giudice, E.~W.~Kolb and A.~Riotto,
  Phys.\ Rev.\ D {\bf 64}, 023508 (2001)
  [hep-ph/0005123]. 

 \bibitem{Gelmini:2006pw} 
  G.~B.~Gelmini and P.~Gondolo,
  Phys.\ Rev.\ D {\bf 74}, 023510 (2006)
  [hep-ph/0602230];
  T.~Moroi and L.~Randall,
  Nucl.\ Phys.\ B {\bf 570}, 455 (2000)
  [hep-ph/9906527];   
  B.~S.~Acharya, P.~Kumar, K.~Bobkov, G.~Kane, J.~Shao and S.~Watson,
  JHEP {\bf 0806}, 064 (2008)
  [arXiv:0804.0863 [hep-ph]];
 P.~Grajek, G.~Kane, D.~Phalen, A.~Pierce and S.~Watson,
  Phys.\ Rev.\ D {\bf 79}, 043506 (2009)
  [arXiv:0812.4555 [hep-ph]];
  B.~Dutta, L.~Leblond and K.~Sinha,
  Phys.\ Rev.\ D {\bf 80}, 035014 (2009)
  [arXiv:0904.3773 [hep-ph]].
  
\bibitem{Erickcek:2011us} 
  A.~L.~Erickcek and K.~Sigurdson,
  Phys.\ Rev.\ D {\bf 84}, 083503 (2011)
\bibitem{Fan:2014zua} 
  J.~Fan, O.~Ozsoy and S.~Watson,
  Phys.\ Rev.\ D {\bf 90}, no. 4, 043536 (2014)
  [arXiv:1405.7373 [hep-ph]].
  
\bibitem{Erickcek:2015jza} 
  A.~L.~Erickcek,
  arXiv:1504.03335 [astro-ph.CO].
  
\bibitem{Lee:2015uza} 
  G.~Lee and C.~E.~M.~Wagner,
  arXiv:1508.00576 [hep-ph];  B.~Li and C.~E.~M.~Wagner,
  Phys.\ Rev.\ D {\bf 91}, 095019 (2015)
  [arXiv:1502.02210 [hep-ph]]; E.~Bagnaschi, G.~F.~Giudice, P.~Slavich and A.~Strumia,
  JHEP {\bf 1409}, 092 (2014)
  [arXiv:1407.4081 [hep-ph]]; A.~Anandakrishnan, B.~Shakya and K.~Sinha,
  Phys.\ Rev.\ D {\bf 91}, 035029 (2015)
  [arXiv:1410.0356 [hep-ph]].
 
\bibitem{Ade:2015xua} 
  P.~A.~R.~Ade {\it et al.} [Planck Collaboration],
  arXiv:1502.01589 [astro-ph.CO].

  
\bibitem{Gelmini:2008sh} 
  G.~B.~Gelmini and P.~Gondolo,
  JCAP {\bf 0810}, 002 (2008)
  [arXiv:0803.2349 [astro-ph]].
  
  
\bibitem{Hofmann:2001bi} 
  S.~Hofmann, D.~J.~Schwarz and H.~Stoecker,
  Phys.\ Rev.\ D {\bf 64}, 083507 (2001)
  [astro-ph/0104173]; 
  T.~Bringmann and S.~Hofmann,
  JCAP {\bf 0704}, 016 (2007)
  [hep-ph/0612238];  S.~Profumo, K.~Sigurdson and M.~Kamionkowski,
  Phys.\ Rev.\ Lett.\  {\bf 97}, 031301 (2006)
  [astro-ph/0603373].
  
  
\bibitem{Gondolo:2004sc} 
  P.~Gondolo, J.~Edsjo, P.~Ullio, L.~Bergstrom, M.~Schelke and E.~A.~Baltz,
  JCAP {\bf 0407}, 008 (2004)
  [astro-ph/0406204].
  
  
  
\bibitem{Press:1973iz} 
  W.~H.~Press and P.~Schechter,
  Astrophys.\ J.\  {\bf 187}, 425 (1974).

  
\bibitem{Ackermann:2015zua} 
  M.~Ackermann {\it et al.} [Fermi-LAT Collaboration],
  arXiv:1503.02641 [astro-ph.HE].


\end{thebibliography}
\end{document}